\documentclass{article}

\usepackage{epsfig}

\date{}

\twocolumn

\begin{document}


\newcommand{\pd}{\partial}
\newcommand{\del}{{\bf \nabla }}
\newcommand{\dv}{{\bf \nabla \cdot}}
\newcommand{\rh}{\overline{\rho}}
\newcommand{\uvr}{{\bf \hat{r}}}
\newcommand{\uvt}{{\bf \hat{\theta}}}


\title{Differential Rotation and Meridional Circulation in Global
Models of Solar Convection\footnote{Published in Astronomische Nachrichten, vol.\ 328, 
no.\ 10, pp.\ 998-1001 (2007)}}

\author{M.S. Miesch (HAO/NCAR, Boulder, CO, 80307-3000; miesch@ucar.edu)} 

\maketitle

\abstract{%
In the outer envelope of the Sun and in other stars, differential rotation and meridional 
circulation are maintained via the redistribution of momentum and energy by convective 
motions.  In order to properly capture such processes in a numerical model, the correct 
spherical geometry is essential.  In this paper I review recent insights into the 
maintenance of mean flows in the solar interior obtained from high-resolution simulations 
of solar convection in rotating spherical shells.   The Coriolis force induces a
Reynolds stress which transports angular momentum equatorward and also yields latitudinal 
variations in the convective heat flux.   Meridional circulations induced by baroclinicity 
and rotational shear further redistribute angular momentum and alter the mean stratification.  
This gives rise to a complex nonlinear interplay between turbulent convection, differential 
rotation, meridional circulation, and the mean specific entropy profile.  I will describe 
how this drama plays out in our simulations as well as in solar and stellar convection zones.}

\section{Introduction}

Axisymmetric flows are a key component of virtually all solar dynamo models.
Differential rotation is thought to be the principal mechanism by which
toroidal field is continually regenerated from poloidal field.
In the language of mean-field dynamo theory, this is known as the 
$\Omega$-effect and the Sun may be classified as an $\alpha$-$\Omega$ dynamo
(e.g.\ Ossendrijver 2003; Charbonneau 2005).  Furthermore, many 
recent dynamo models, known as flux-transport models, suggest that 
global axisymmetric circulations in the meridional plane (latitude-radius)
may account for the observed migration of active regions toward the equator 
during the course of the solar activity cycle and may thus
determine the cycle period (e.g.\ Wang et al. 1991; Choudhuri et al.\ 1995; 
Dikpati \& Charbonneau 1999; Dikpati \& Gilman 2006).

The internal rotation profile of the Sun is well 
established from helioseismology (Thompson et al. 2003)
and as we have seen at this meeting, there is ample
evidence for the presence of differential rotation
in other stars (see other papers in these proceedings
as well as previous work by Barnes et al.\ 2005 and
Reiners 2006).  In the solar convection zone the angular 
velocity decreases monotonically by about 30\% from 
equator to pole with nearly radial contours at mid 
latitudes.  Regions of stronger radial shear exist near
the top and bottom of the convective zone, the latter 
known as the solar tachocline.

The meridional circulation is more challenging to detect
but photospheric measurements and local helioseismic
inversions have revealed a systematic but variable 
poleward flow of 15-20 m s$^{-1}$ in the upper 
solar convection zone (e.g. Hathaway 1996; 
Haber et al.\ 2002; Gonz\'ales-Hern\'andez et al.\ 2006).

\begin{figure*}
\epsfig{file=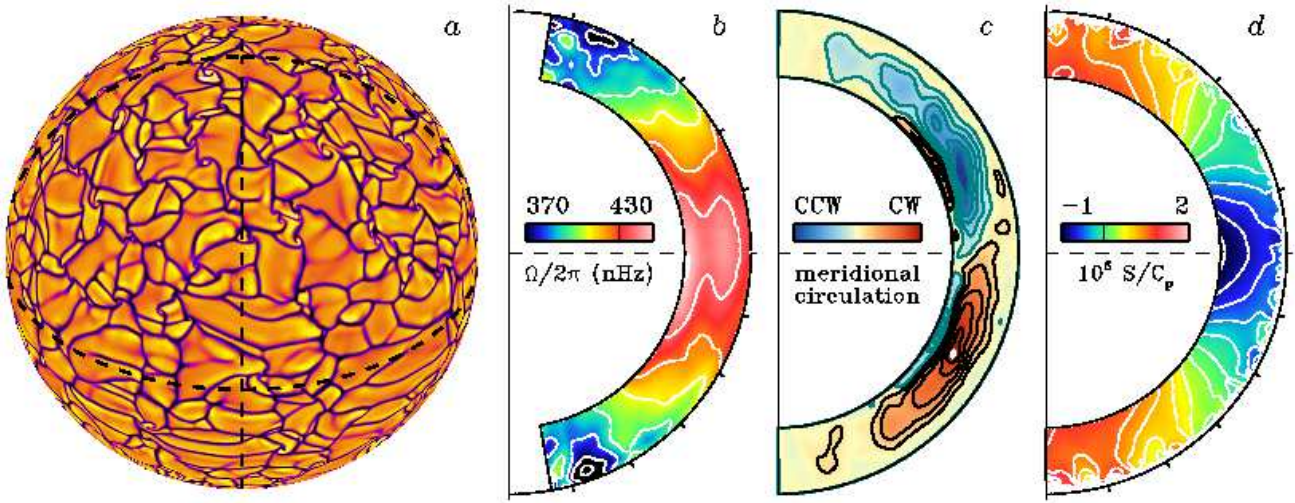,width=7.0in}
\caption{Flow features in a simulation of solar convection.
($a$) Radial velocity near the top of the computational 
domain ($r = 0.98R$) with yellow tones denoting upflow and
blue/black tones  denoting downflow. This orthographic 
projection is tilted 35$^\circ$ toward the line of sight
so the north pole is visible.  ($b$)
Differential rotation, ($c$) meridional circulation, and
($d$) specific entropy perturbations averaged over longitude
and time (132 days).  The meridional circulation
is represented as streamlines of the mass flux and
the entropy is normalized by the specific heat
at constant pressure $C_P = 3.5 \times 10^8$
erg g$^{-1}$ K$^{-1}$ (after Miesch et al.\ 2007).}
\end{figure*}

Differential rotation and meridional circulation in the solar 
envelope and in other stars are maintained by convection.
The highly turbulent nature of solar and stellar convection
makes numerical modeling a difficult challenge but contininuing
advances in supercomputing technology have enabled increasingly
realistic 3D simulations.  In this paper I'll review what our
most recent convection simulations reveal about the nature
of mean flows in the solar envelope and how they are maintained.

\section{Turbulent Solar Convection}\label{convection}

High-resolution simulations of solar convection exhibit
intricate flow structures and persistent mean flows as
illustrated in Figure 1.  The simulation shown was carried 
out using the ASH (Anelastic Spherical Harmonic) computer
code described by Clune et al.\ (1999) and Brun, Miesch \& 
Toomre (2004).  The ASH code solves the three-dimensional
equations of hydrodynamics (or magneto-hydrodynamics) in
a rotating spherical-shell geometry under the anelastic
approximation.  The simulation shown in Figure 1 extends
from the base of the convective envelope at $0.71R$, where
$R$ is the solar radius, to $0.98R$, within about 14 Mm
of the photosphere.  Beyond $0.98R$, granulation and 
supergranulation occur in the Sun which cannot be 
resolved in a global model (yet) and which involve physical 
processes neglected in ASH such as ionization and radiative 
transfer.  Solar values are used for the luminosity and rotation 
rate and the background stratification is based on a solar
structure model.  For further details on this simulation
see Miesch et al.\ (2007).

Near the top of the computational domain there is an 
interconnected network of downflow lanes which appears
similar to solar granulation but occupies a much larger 
scale (Fig. 1$a$).  These {\em giant cells} are about 
100 Mm across on average, compared to 30 Mm for 
supergranulation and 1 Mm for granulation.  At high 
and mid latitudes, the downflow lanes exhibit intense 
cyclonic vorticity (counter-clockwise in the northern 
hemisphere, clockwise in the southern), which is evident 
in animations of the convective patterns.

The correlation time for the downflow network is 
2-3 days, depending on latitude, but closer scrutiny
reveals longer-lived structures (Miesch et al.\ 2007).  
At low latitudes, persistent downflow lanes occur 
which are oriented in a north-south direction and 
which persist for weeks to months.  These are traveling 
convection modes which propagate prograde faster than 
the local rotation rate.  Although the downflow network
fragments deeper down, the north-south downflow
lanes extend through most of the convection zone
and play an important role in the maintenence of
mean flows (\S3).

The differential rotation achieved in this simulation
is shown in Figure 1$b$.  It is roughly solar-like, 
with a monotonic decrease in the angular velocity 
from equator to pole and relatively little variation
with radius across the convection zone.  However,
the angular velocity contrast between the equator
and $\pm 60^\circ$ is only about 50 nHz, compared
to 90 nHz in the Sun.  Other ASH simulations have
achieved better agreement with helioseismic inversions,
including a stronger angular velocity contrast and
nearly radial angular velocity contours at mid 
latitudes (Miesch et al.\ 2006).  

Figure 1$c$ shows the meridional circulation which
is dominated by a single cell in each hemisphere, with 
poleward flow in the upper convection zone 
(amplitude 15-20 m s$^{-1}$) and equatorward flow 
in the lower convection zone (amplitude 5-10 m s$^{-1}$).  
This is roughly consistent with flux transport solar
dynamo models and helioseismic determinations of the 
meridional circulation in the upper convection
zone of the Sun (\S1).

Although the circulation pattern in Figure 1$c$ is
predominatly single-celled in each hemisphere,
narrow counter-cells occur near the top and 
bottom boundaries.  In light of our idealized
boundary conditions (stress-free, impenetrable,
imposed specific entropy profile on the bottom and
constant heat flux on the top), the presence of these 
counter-cells should be interpreted with caution.
ASH simulations which incorporate convective
penetration into an underlying radiative zone
generally exhibit equatorward meridional 
circulation throughout the overshoot region
(Miesch et al.\ 2000).  Furthermore, angular 
momentum transport by supergranulation may
induce a systematic poleward circulation in
the surface layers via a process known as 
gyroscopic pumping (see eq.\ (1) below).

Figure 1$a$, $b$, and $c$ illustrate convection, differential
rotation, and meridional circulation.  Before proceeding to
\S3, there is one more player in this drama to be introduced.
Figure 1$d$ shows the mean specific entropy perturbation
$\left<S\right>$ relative to the spherically symmetric 
background stratification, averaged over longitude and time.
The $\left<S\right>$ profile exhibits prominent {\em warm poles}
associated with thermal wind balance (\S3).  The corresponding
temperature variation is $\sim $ 8K (Miesch et al.\ 2007, Fig.\ 6)
which is more than five orders of magnitude smaller than the
background temperature of 2.2 million K at the base of the 
convection zone.  However, whereas the specific entropy 
perturbation increases monotonically with latitude throughout 
the convection zone, the temperature perturbation near the 
top of the shell typically peaks at the poles and the equator, 
with a minimum at mid-latitudes (Brun \& Toomre 2002; 
Miesch et al.\ 2006).

\begin{figure}
\epsfig{file=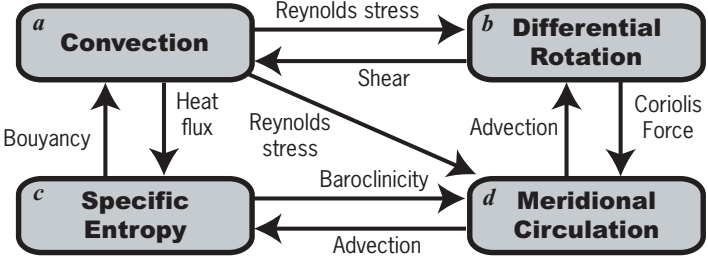,width=\linewidth}
\caption{Schematic diagram illustrating the complex nonlinear 
coupling between ($a$) convection,  ($b$) differential rotation, 
($c$) the mean specific entropy profile $\left<S\right>$,
and ($d$) the meridional circulation (see text).}
\end{figure}

\section{Maintenance of Mean Flows}\label{meanflows}

In order to understand how mean flows are maintained in 
our convection simulations and, by extension, in the
Sun and in other stars, two equations are particularly
enlightening.  The first is derived from the zonal (longitudinal)
component of the momentum equation, averaged over 
longitude and time:
\begin{equation}\label{amom}
\dv \left(\rh \left<{\bf v}_m\right> {\cal L}\right) = 
- \dv \left(\rh r \sin\theta \left<v_\phi^\prime {\bf v}_m^\prime\right>\right) 
\end{equation}
where ${\cal L}$ is the specific angular momentum
\begin{displaymath}
{\cal L} = r \sin\theta \left(\Omega r \sin\theta + \left<v_\phi\right>\right) ~~~.
\end{displaymath}
Equation (\ref{amom}) is expressed in a rotating spherical polar coordinate 
system with radius $r$, colatitude $\theta$, and longitude $\phi$ and 
corresponding velocity components $v_r$, $v_\theta$, and $v_\phi$.  
The meridional velocity component is defined as 
${\bf v}_m = v_r \uvr + v_\theta \uvt$ where $\uvr$ and
$\uvt$ are unit vectors in the $r$ and $\theta$ directions.  Brackets
$<>$ denote averages over longitude and time, and primes indicate
fluctuating variables for which the mean has been subtracted out,
e.g.\ $v_\phi^\prime = v_\phi - \left<v_\phi\right>$.  The rotation
rate of the coordinate system is $\Omega$ and $\rh$ is the
(spherically symmetric) background density stratification.

The second equation of particular importance is derived from
the zonal component of the vorticity equation which is
the curl of the momentum equation:
\begin{equation}\label{tw}
{\bf \Omega \cdot \nabla} \left<v_\phi\right> = 
\frac{g}{2 r C_P} \frac{\pd \left<S\right>}{\pd \theta} 
\end{equation}
where $\left<S\right>$ is the mean specific entropy 
perturbation as in Fig.\ 1$d$, $g$ is the gravitational
acceleration, and $C_P$ is the specific heat per unit
mass at constant pressure.

When deriving equations (1) and (2) we have assumed that 
the system in question (whether it be a simulation or 
a star) is in a statistically steady state.  Furthermore,
we have neglected viscous
dissipation which is significant in some simulations but
insignificant in stellar interiors where the Reynolds
number (the ratio of inertial to viscous forces) is
of order 10$^{12}$ or more.  We have also neglected
the Lorentz force which generally has a dissipative
effect in the convection zone, suppressing rotational 
shear (Brun et al.\ 2004).  In the tachocline, Lorentz 
forces associated with strong, localized toroidal field 
structures may divert or induce meridional and zonal flows.

Equation (2) involves several additional assumptions,
most notably that the Coriolis force operating on the 
differential rotation is large relative to the
Reynolds stress.  In other words, we have assumed
that the Rossby number $Ro = \omega / 2\Omega$ is
much less than unity, where $\omega$ is the fluid 
vorticity in the rotating frame.  The Rossby number
is not specified in ASH simulations; rather, it is
computed {\em a posteriori} from the simulated
flow field.  Results generally indicate that the
small $Ro$ assumption implied by equation (2) is
valid in the lower convection zone but breaks 
down near the surface where the amplitude of
$\omega$ peaks.

When deriving equation (2) we have also assumed 
an ideal gas equation of state and that the background 
stratification is hydrostatic and adiabatic to lowest order.
Both of these assumptions are well justified in
the bulk of the solar convection zone but break down
for $r > 0.98 R$.

Subject to these reasonable assumptions, equation (1)
states that the advection of angular momentum by
the meridional circulation $\left<{\bf v}_m\right>$ must 
balance the angular momentum transport due to the 
Reynolds stress $\left<v_\phi^\prime {\bf v}_m^\prime\right>$.
The Reynolds stress arises from correlations in the
convective velocity components induced by the
Coriolis force.  Of particular relevance are the
north-south oriented downflow lanes mentioned 
in \S2; as horizontal flows near the top of the
convection zone converge into these downflow
lanes, eastward flows are diverted equatorward
by the Coriolis force and westward flows are 
diverted poleward.  This produces an equatorward
angular momentum transport which is mainly
responsible for the prograde equatorial rotation 
seen in Figure 1$b$.

Equation (2) states that gradients in the mean zonal
velocity (as well as the mean angular velocity)
parallel to the rotation axis are proportional to
the latitudinal entropy gradient.  This is the solar
analogue of thermal wind balance which has been 
studied for decades within the context of geophysics
(e.g.\ Pedlosky 1987).  In the absence of latitudinal
entropy gradients, equation (2) implies cylindrical
rotation profiles in which angular velocity (and zonal
velocity) contours are parallel to the rotation axis.
This is a manifestation of the well-known Taylor-Proudman
theorem.

Insight into how the dynamical balances expressed by
equations (1) and (2) are achieved can be gained by 
considering Figure 2.  Convection redistributes momentum
and can establish differential rotation (DR) and meridional
circulation (MC) through the Reynolds stress ($a \rightarrow b$,
$a \rightarrow d$).  It can also
alter the mean entropy profile by means of the convective
heat flux which includes contributions from enthalpy flux
as well as kinetic energy flux ($a \rightarrow c$).  
Differential rotation can
provide a negative feedback by shearing out convection
cells ($b \rightarrow a$).  Meanwhile, modications of the 
specific entropy profile alter the buoyancy driving of the 
convective motions ($c \rightarrow a$).  The meridional 
circulation can in principle feed back directly on the 
convection as well ($d \rightarrow a$) but in 
practice this is negligible because the convective kinetic
energy generally exceeds that contained in the MC by 
about two orders of magnitude.

Differential rotation can induce meridional circulations \\
through the Coriolis force ($b \rightarrow d$) while entropy variations
can induce circulations through baroclinicity ($c \rightarrow d$), in 
particular the $\pd \left<S\right>/\pd \theta$ term
on the right-hand-side of equation (2).  Advection
of angular momentum and entropy by the meridional 
circulation can in turn alter the $\left<v_\phi\right>$
and the $\left<S\right>$ profiles ($d \rightarrow b$,
$d \rightarrow c$).

Equation (1) involves panels $a$, $b$, and $d$ of
Figure 2, such that angular momentum transport
by the convective Reynolds stress ($a \rightarrow b$) 
balances that by the meridional circulation
($d \rightarrow b$).  It is this balance that
largely determines the meridional circulation
pattern.  As the Reynolds stress establishes a differential
rotation, the meridional circulation adjusts such
that equation (1) is satisfied.  In previous, more
laminar simulations of convection, angular momentum
transport by viscous diffusion upset the balance
expressed by equation (1) and circulation patterns
were qualitatively different from that in Figure 1$c$, 
multi-celled in latitude and radius (Miesch et al.\ 2000, 2006; 
Brun \& Tooomre 2002).

This simplified picture is complicated by the nonlinear
feedback mechanisms illustrated in Figure 2 and the
additional constraint provided by thermal wind
balance, equation (2).  Equation (2) only explicitly
involves the differential rotation and specific 
entropy profile ($b$, $c$), but convection and
meridional circulation ($a$, $d$) both play an essential
role.  In order to demonstrate this, we consider now 
two illustrative scenarios by which thermal wind balance
may be achieved.  We'll refer to these as mechanical
and thermal forcing (Miesch et al.\ 2006).  Both cases
begin with convection ($a$) but they each circle the 
diagram in Figure 2 in an opposite sense, clockwise 
and counter-clockwise respectively.

In the mechanical forcing scenario, the convective
Reynolds stress establishes a differential 
rotation ($a \rightarrow b$) which induces a meridional 
circulation via the Coriolis force ($b \rightarrow c$).  
The advection of entropy by these circulations then 
alters the background stratification $\left<S\right>$ 
($c \rightarrow d$).  In the subadiabatic portion of 
the tachocline, this would tend to establish warm poles 
and the Coriolis-induced circulations would cease when 
thermal wind balance, equation (2), is achieved 
(Rempel 2005; Miesch et al 2006).  However, in the 
convection zone the sense of the induced circulation is 
such that the resulting latitudinal entropy gradient 
would be equatorward 
rather than poleward (cool poles).  Thermal wind balance 
would be achieved eventually, but the rotation profile 
would not be solar-like.

Now consider thermal forcing in which a latitude-dependent
convective heat flux establishes warm poles ($a \rightarrow c$)
which in turn induce a meridional circulation through
baroclinicity ($c \rightarrow d$).  Redistribution of
angular momentum by this baroclinic circulation then
establishes a differential rotation ($d \rightarrow b$).
This will `work' in the sense that the induced DR will 
evolve until equation (2) is satisfied.  However, 
in the absence of the Reynolds stress, the meridional
circulation would tend to conserve angular momentum,
accelerating the poles relative to the equator.
This follows from equation (1); if the right-hand-side
is zero and $\dv \left(\rh \left<{\bf v}_m\right>\right) = 0$
as required by the anelastic approximation, then 
$\left<{\bf v}_m\right> {\bf \cdot \nabla} {\cal L} = 0$,
so ${\cal L}$ would be constant along circulation streamlines.
A solar-like differential rotation profile can only
be achieved if the Reynolds stress also contributes,
transporting angular momentum toward the 
equator ($a \rightarrow b$).

Thus, the convective Reynolds stress is needed to produce
a prograde equatorial rotation relative to higher latitudes
as in the Sun while baroclinicity is needed to break the
Taylor-Proudman preference for cylindrical rotation
profiles.  Such baroclinicity arises from a latitude-dependent 
convective heat flux but thermal coupling to the tachocline 
may also play an important role (Rempel 2005; Miesch et al.\ 2006).
Much has been learned but much remains to be done and convection
simulations will continue to provide valuable insight into how
differential rotation and meridional circulation are established
and maintained in the Sun and in stars across the HR diagram.

\vspace{.1in}
The work reported here is done in close collaboration with our ``ASH mob''
of Allan Sacha Brun, Juri Toomre, Marc DeRosa, Matthew Browning, 
Benjamin Brown, Nicholas Featherstone, Kyle Augustson, and Nicholas
Nelson.  Funding was provided by NASA through grant NNG05G124G of the
Heliophysics Theory Program.  The simulations were carried out with
NSF PACI support of PSC, SDSC, and NCSA and NASA support through
Project Columbia.


\end{document}